\documentclass[conference]{IEEEtran}
\IEEEoverridecommandlockouts
\usepackage{cite}
\usepackage{amsmath,amssymb,amsfonts}
\usepackage{algorithm} 
\usepackage{graphicx}
\usepackage{textcomp}
\usepackage{xcolor}
\usepackage{mathtools}
\usepackage{algorithmic}
\usepackage{hyperref}
\usepackage{booktabs}
\usepackage{balance}
\usepackage{bbm}
\usepackage{float}

\def\BibTeX{{\rm B\kern-.05em{\sc i\kern-.025em b}\kern-.08em
    T\kern-.1667em\lower.7ex\hbox{E}\kern-.125emX}}
\begin{document}

\title{A Variational Bayesian Detector for Affine Frequency Division Multiplexing\\}
\author{Can Zheng\\
School of Electrical Engineering, Korea University, Seoul, Republic of Korea\\
Email: \{zc331\_\}@korea.ac.kr\\
}

\author{Can Zheng and Chung G. Kang,~\IEEEmembership{Senior Member,~IEEE}

\thanks{(Corresponding author: Chung G. Kang.)}
\thanks{This work was supported by the National Research Foundation of Korea (NRF) grant funded by the Korea government (MSIT) (RS-2025-00517140).
Can Zheng and Chung Gu Kang are with the School of Electrical Engineering, Korea University, Seoul 02841, South Korea (e-mail: zc331\underline{~}@korea.ac.kr; ccgkang@korea.ac.kr).}}
\maketitle
\begin{abstract}
    This paper proposes a variational Bayesian (VB) detector for affine frequency division multiplexing (AFDM) systems. The proposed method estimates the symbol probability distribution by minimizing the Kullback-Leibler (KL) divergence between the true posterior and an approximate distribution, thereby enabling low-complexity soft-decision detection. Compared to conventional approaches such as zero-forcing (ZF), Linear minimum mean square rrror (LMMSE), and the message passing algorithm (MPA), the proposed detector demonstrates lower bit error rates (BER), faster convergence, and improved robustness under complex multipath channels. Simulation results confirm its dual advantages in computational efficiency and detection performance.
\end{abstract}


\section{Introduction}

    Reliable wireless communication in high-mobility scenarios is critical for next-generation networks. Orthogonal frequency division multiplexing (OFDM), which is the dominant waveform in 4G and 5G, excels in time-invariant frequency-selective channels but degrades in high-speed environments like vehicular networks and LEO satellite communications due to Doppler-induced inter-carrier interference (ICI). To overcome these limitations, orthogonal time frequency space (OTFS) modulation has been proposed \cite{OTFS}. Operating in the delay-Doppler (DD) domain, OTFS exploits the quasi-static and sparse nature of wireless channels to mitigate ICI \cite{OTFS3}. However, its two-dimensional modulation structure introduces notable computational complexity. In contrast, affine frequency division multiplexing (AFDM) utilizes a more efficient one-dimensional structure that maintains full diversity with reduced complexity \cite{AFDM}. As a relatively new waveform, AFDM still faces open problems, particularly in channel estimation and equalization under time-varying conditions.

    Detection algorithms for AFDM systems present fundamental trade-offs. Maximum \textit{a posteriori} (MAP) detection is optimal but computationally infeasible. Linear detectors, such as zero-forcing (ZF) and linear minimum mean square error (LMMSE), reduce complexity through matrix inversion but still incur significant computational costs. Additionally, ZF suffers from noise enhancement, while LMMSE, though more robust, remains suboptimal in doubly-dispersive environments. Message passing algorithm (MPA)-based iterative detectors exploit channel sparsity for improved efficiency and have been widely applied in wireless communication systems, including low-density parity-check (LDPC) codes and sparse code multiple access (SCMA), etc \cite{MPA2, MPA_LDPC, MPA_SCMA, MPA}. However, MPA faces limitations in multipath channels, where loops in the factor graph can lead to suboptimal solutions or slow convergence. 

    This paper proposes a variational Bayesian (VB) detector for AFDM systems \cite{PRML, VB}, which approximates the posterior distribution by minimizing the Kullback-Leibler (KL) divergence. Unlike MPA, the VB detector ensures global consistency, faster convergence, and reduced complexity. Simulation results show superior bit error rate (BER) performance and practical feasibility. The remainder of this paper is organized as follows: Section II presents the AFDM system model. Section III describes the proposed algorithm. Section IV presents and discusses simulation results. Section V concludes the paper.

    \textit{Notations}: Bold lowercase letters denote vectors (e.g., $\mathbf{x}$), and bold uppercase letters denote matrices (e.g., $\mathbf{X}$). The superscripts $(\cdot)^\mathsf{H}$ represent the Hermitian (conjugate transpose) operation. $\propto$ denotes proportionality
    and $\triangleq$ indicates a definition. The operator $\mathbb{E}[\cdot]$ denotes the statistical expectation, while $|\cdot|_2$ denotes the Euclidean norm of a vector, and $|\cdot|$ returns the magnitude of a complex number. $\delta(\cdot)$ represents the Dirac delta function. $\mathcal{O}(\cdot)$ is an asymptotic notation denotes the order of computational complexity. The complex Gaussian distribution is denoted by $\mathcal{CN}(\mu,\sigma^2)$ with mean $\mu$ and variance $\sigma^2$. Unless otherwise specified, $\mathbb{Z}$, $\mathbb{R}$, and $\mathbb{C}$ denote the sets of integer, real, and complex numbers, respectively.
    
\section{System Model}
    \subsection{AFDM Modulation and Demodulation Framework}
    Let $\mathbb{A} = \{a_1, \ldots, a_K\}$ denote a $K$-ary quadrature amplitude modulation (QAM) constellation defined over the Gaussian integer ring $\mathbb{Z}[j]$, where each constellation point satisfies $z_r + jz_i$ with $z_r$, $z_i \in \mathbb
    Z$. The time-domain modulated symbols $\{s_n\}_{n=0}^{N-1}$ is generated through the inverse discrete affine Fourier transform (IDAFT) of the information symbols $\{x_m\}_{m=0}^{N-1}$ as follows:
    \begin{align}
    s_n = \frac{1}{\sqrt{N}} \sum_{m=0}^{N-1} x_m e^{j2\pi \left(c_1 n^2 + c_2 m^2 + \frac{nm}{\sqrt{N}}\right)},
    \label{time_transmit}
    \end{align}
    for $n = 0,\ldots,N-1$, where $c_1$, $c_2\in \mathbb{R}$ are parameters specific to the AFDM system, optimized based on the DD characteristics of the channel. To mitigate inter-symbol interference (ISI), the following chirp-periodic prefix (CPP) is appended to the modulated signal:
    \begin{align}
        s_n = s_{N+n} e^{j2\pi c_1(N^2 +2Nn)}, n=-L_\text{CPP},\cdots, -1,
    \end{align}
    where $L_\text{CPP}$ is any integer greater than or equal to the number of samples corresponding to the maximum delay spread of the channel. The CPP reduces to a conventional CP when $2Nc_1\in \mathbb{Z}$ and $N$ is even. 
    
    The time-domain received signal after transmission over a time-varying multipath channel is expressed as
        \begin{align}
        r_n = \sum_{l=0}^\infty s_{n-l}g_{n,l} +w_n,\label{time receive}
    \end{align}
    where $w_n \sim \mathcal{CN}(0,N_0)$ is the additive white Gaussian noise (AWGN) and the channel impulse response $g_{n,l}$ at time $n$ and delay $l$ is given by
    \begin{align}
        g_{n,l} = \sum_{i=1}^{P} h_i e^{j2\pi \nu_i \frac{n}{N}} \delta(l-l_i),
    \end{align}
    with $P$ denoting the number of paths, and $h_i$, $f_i$ and $l_i$ representing the complex gain, normalized Doppler shift, and integer delay of the $i$-th path, respectively. 
    
    After removing the CPP, the received signal is demodulated via DAFT as follows:
    \begin{align}
        y_m = \sum_{n=0}^{N-1}r_n e^{-j2\pi \left(c_1n^2 +c_2 m^2 +\frac{nm}{\sqrt{N}}\right)}
    \end{align}
    for $m = 0,\cdots, N-1$.
    \subsection{Matrix Form of Input-Output Relationship}
    Given the information symbol vector $\mathbf{x}\in \mathbb{A}^{N\times 1}$ and the transformation (\ref{time_transmit}), the transmitted signal vector $\mathbf{s} \in \mathbb{C}^{N \times 1}$ can be expressed as
    \begin{align}
        \mathbf{s} = \mathbf{A}^\mathsf{H} \mathbf{x}
        =\mathbf{\Lambda}_{c_1}^\mathsf{H} \mathbf{F}^\mathsf{H} \mathbf{\Lambda}_{c_2}^\mathsf{H} \mathbf{x}, \label{transmit matrix}
    \end{align}
    where $\mathbf{A}=\mathbf{\Lambda}_{c_2}\mathbf{F}\mathbf{\Lambda}_{c_1}$ is, $\mathbf{F}$ is the discrete Fourier transform (DFT) matrix with elements $[\mathbf{F}]_{n,m}=\frac{1}{\sqrt{N}}e^{-j2 \pi \frac{nm}{N}}$, and $\mathbf{\Lambda}_{c}=\text{diag}(e^{-j2 \pi c n^2}, n=0,\cdots, N-1)$. For the case of integer delays, the time-domain channel matrix $\mathbf{H}\in \mathbb{C}^{N \times N}$ incorporating the CPP can be expressed as
    \begin{align}
        \mathbf{H} = \sum_{p=1}^P h_p \mathbf{\Gamma}_{\text{CPP}_p}\mathbf{\Delta}_{f_p}\mathbf{\Pi}^{l_p},
    \end{align}
    where $\mathbf{\Pi}$ is the permutation matrix, i.e.,
    \begin{align}
        \mathbf{\Pi} =
        \begin{bmatrix}
            0 & 0 & \cdots & 0 & 1 \\
            1 & 0 & \cdots & 0 & 0 \\
            0 & 1 & \cdots & 0 & 0 \\
            \vdots & \vdots & \ddots & \vdots & \vdots \\
            0 & 0 & \cdots & 1 & 0 \\
        \end{bmatrix}_{N\times N},
    \end{align}
    $\mathbf{\Delta}_{f_p}$ and $\mathbf{\Gamma}_{\text{CPP}_p}$ are the following diagonal matrices:
    \begin{align}
        \mathbf{\Delta}_{f_p} &= \text{diag}\left(e^{-j2\pi \frac{\nu_p n}{N}}\right), \\
        \mathbf{\Gamma}_{\text{CPP}_p} &= \text{diag}\left(
        \begin{cases} 
            1, & \text{if } n \ge l_p, \\
            e^{-j2\pi c_1 (N^2 + 2N(n-l_p))}, & \text{if } n<l_p
        \end{cases}\right).
    \end{align}
    The matrix $\mathbf{\Gamma}_{\text{CPP}_p}$ compensates for the phase rotation introduced by the CPP, ensuring the periodicity of the time-domain signal for efficient channel estimation and demodulation. According to (\ref{time receive}), the received symbol vector $\mathbf{r}$ is given by
    \begin{align}
        \mathbf{r} = \mathbf{H}\mathbf{s}+\mathbf{w}, \label{receive matrix1}
    \end{align}
    where $\mathbf{w}\sim \mathcal{CN}(\mathbf{0},N_0\mathbf{I})$ is the noise vector. Finally, after DAFT, the received signal in
    the DAF domain is given by:
    \begin{align}
        \mathbf{y} &= \mathbf{A}\mathbf{r}. \label{receive matrix2}
    \end{align}
Substituting (\ref{transmit matrix}) and (\ref{receive matrix1}) into (\ref{receive matrix2}), the input-output relationship can then be expressed as
    \begin{align}
        \mathbf{y} &= \mathbf{A}\mathbf{r} = \mathbf{A}\mathbf{H}\mathbf{s}+\mathbf{A}\mathbf{w}\nonumber \\
        &=\mathbf{A}\mathbf{H}\left(\mathbf{A}^\mathsf{H} \mathbf{x} \right) + \mathbf{A}\mathbf{w} \nonumber \\
        &=\bar{\mathbf{H}} \mathbf{x}+\bar{\mathbf{w}}, \label{i/o}
    \end{align}
    where $\bar{\mathbf{H}}\triangleq \mathbf{A}\mathbf{H}\mathbf{A}^\mathsf{H}$ is the effective channel matrix, and $\bar{\mathbf{w}
    }\triangleq \mathbf{A}\mathbf{w}$ is the effective noise.
    
\section{Detector Design}
    \subsection{Variational Inference Framework}
    The objective of signal detection is to recover the transmitted signal vector $\mathbf{x}$ from the received signal $\mathbf{y}$. This problem is formulated within a Bayesian framework as a MAP estimation task\cite{Digital_Comm}, i.e., 
    \begin{align}
        \hat{\mathbf{x}}_\text{MAP} 
        &= \arg \max_{\mathbf{x} \in \mathbb{A}^{N\times 1}} p(\mathbf{x}|\mathbf{y}).
        \label{eq:MAP}
    \end{align}
    According to Bayes' rule, the \emph{a posteriori} probability of $\mathbf{x}$ can be expressed as
    \begin{align}
        p(\mathbf{x}|\mathbf{y}) = \frac{p(\mathbf{x})p(\mathbf{\mathbf{y}|x})}{p(\mathbf{y})} \propto p(\mathbf{x})p(\mathbf{\mathbf{y}|x}),
    \end{align}
    where the likelihood function follows a complex Gaussian distribution, i.e., 
    \begin{align}
        p(\mathbf{\mathbf{y}|x}) = \mathcal{CN} (\mathbf{y}|\bar{\mathbf{H}}\mathbf{x}, N_0 \mathbf{I}) = \frac{1}{(\pi N_0)^N}e^{-\frac{\Vert \mathbf{y}-\bar{\mathbf{H}}\mathbf{x}\Vert^2 }{N_0}},\label{eq:likelihood}
    \end{align}
    and the prior distribution is uniform, as it is assumed that the transmitted symbols are uniformly distributed over $\mathbb{A}$, i.e., 
    \begin{align}
        p(\mathbf{x})=\prod_{n=1}^Np(x_n) = \prod_{n=1}^N\frac{1}{K}\sum_{k=1}^K \delta(x_n-a_k).\label{eq:prior}
    \end{align}
    
    Direct computation of (\ref{eq:MAP}) via exhaustive search is computationally infeasible due to the exponential growth of possible configurations (i.e., $K^N$ combinations). To address this, we we employ variational inference (VI) to approximate the posterior $p(\mathbf{x}|\mathbf{y})$ with a variational distribution $q(\mathbf{x})$. The optimal variational distribution is obtained by minimizing the KL divergence as follows:
    \begin{align}
        q^*(\mathbf{x}) 
        &= \arg \min_{q \in \mathcal{Q}} \text{KL}\big(q(\mathbf{x}) \Vert p(\mathbf{x}|\mathbf{y})\big) \nonumber \\
        &= \arg \min_{q \in \mathcal{Q}} \mathbb{E}_{q} \left[\log \frac{q(\mathbf{x})}{p(\mathbf{x}|\mathbf{y})}\right],
        \label{eq:KL_min}
    \end{align}
    where $\mathcal{Q}$ is a family of tractable distributions.
    Expanding \eqref{eq:KL_min}, we obtain
    \begin{align}
        \text{KL}(q \Vert p) 
        &=\mathbb{E}_q[\log q(\mathbf{x})] - \mathbb{E}_q[\log p(\mathbf{x}| \mathbf{y})] \nonumber \\
        &=-\underbrace{\left(\mathbb{E}_q[\log q(\mathbf{x})] + \mathbb{E}_q[\log p(\mathbf{x}, \mathbf{y})]\right)}_{\text{ELBO, computable terms}} + \log p(\mathbf{y}).
        \label{eq:KL_expansion}
    \end{align}
        Since $\log p(\mathbf{y})$ is independent of $q$, minimizing $\text{KL}(q \Vert p)$ is equivalent to maximizing the evidence lower bound (ELBO):
    \begin{align}
        \mathcal{L}(q) 
        &= \mathbb{E}_q[\log p(\mathbf{x}, \mathbf{y})] - \mathbb{E}_q[\log q(\mathbf{x})]\nonumber \\
        &= \mathbb{E}_q[\log p(\mathbf{y}| \mathbf{x})] +\mathbb{E}_q[\log p(\mathbf{x})]- \mathbb{E}_q[\log q(\mathbf{x})] \nonumber \\
        &= \mathbb{E}_q[\log p(\mathbf{y}| \mathbf{x})]-\text{KL}(q(\mathbf{x})\Vert p(\mathbf{x})).
        \label{eq:ELBO}
    \end{align}
    
    This equivalence transforms the inference problem into an optimization task over tractable distributions $q(\mathbf{x})$. To make the optimization feasible, we impose the \emph{mean-field assumption}, which factorizes the joint variational distribution into independent marginal distributions as follows:
    \begin{align}
        q(\mathbf{x}) 
        &= \prod_{n=1}^N q_n(x_n),
        \label{eq:mean_field}
    \end{align}
    where each $q_n(x_n) \sim \mathcal{CN}(x_n;\hat{x}_n,\hat{v_n})$ Thus, (\ref{eq:ELBO}) can be rewritten as
    \begin{align}
        \mathcal{L}(q) &= \mathbb{E}_q[\log p(\mathbf{y}| \mathbf{x})]-\sum_{n=1}^N \text{KL}(q_n(x_n)\Vert p(x_n)).
    \end{align}

    We first handle the likelihood term associated with other symbols. According to (\ref{eq:likelihood}),
    \begin{align}
       \mathbb{E}_q\left[\log p(\mathbf{y}|\mathbf{x})\right] &= -\frac{1}{N_0}\mathbb{E}_q\left[\Vert \mathbf{y-\bar{\mathbf{H}} }\mathbf{x}\Vert^2\right] - \mathbb{E}_q \left[N\log(\pi N_0)\right] \nonumber \\
       &= -\frac{1}{N_0} \left(\Vert \mathbf{y}- \bar{\mathbf{H}} \hat{\mathbf{x}} \Vert^2 + \sum_{n=1}^N \Vert \bar{\mathbf{H}}(:.n)\Vert^2 \hat{v}^2_n\right).\label{eq:likelihood2}
    \end{align}
    And then, for each symbol $x_n$,
    \begin{align}
        \text{KL}\left(q_n(x_n) \Vert p(x_n)\right) &= \sum_{k=1}^K \pi^q_n \log \frac{\pi^q_j}{1/K}\nonumber \\ &= \sum_{k=1}^K \pi^k_n \log \pi^k_n + \log K,
    \end{align}
    where $\pi_n^k = q_n(x_n = a_k) = \int \delta(x_n - a_k) q_n(x_n) \, dx_n$ represents the probability that $x_n = a_k$ under the variational distribution.

    By collecting the terms, the variational objective for updating $q_n(x_n)$ is
    \begin{align} \mathcal{L}(q_n) &= -\frac{1}{N_0} \left(\Vert \mathbf{y}- \bar{\mathbf{H}} \hat{\mathbf{x}} \Vert^2 + \sum_{n=1}^N \Vert \bar{\mathbf{H}}(:.n)\Vert^2 \hat{v}^2_n\right) \nonumber \\
    &-\sum_{n=1}^N\left(\sum_{k=1}^K \pi^k_n \log \pi^k_n + \log K\right),
    \end{align}
    where $\bar{\mathbf{H}}{(:,n)}$ denotes the $n$-th column of $\bar{\mathbf{H}}$.

    \subsection{Coordinate Ascent Algorithm}

    To maximize the joint ELBO, we employ the coordinate ascent variational inference (CAVI), iteratively updating each variational factor $q_n(x_n)$ while holding the others fixed. The ELBO for the $n$-th factor is given as 
    \begin{align} 
        \mathcal{L}(q_n) &= \mathbb{E}_{q_j}\left[\mathbb{E}_{q_{\setminus n}}\left[\log p(\mathbf{y}|\mathbf{x})\right]\right] - \text{KL}\left(q_n(x_n)\Vert p_n(x_n)\right)\nonumber \\
        &=-\frac{1}{N_0}\left(\Vert \mathbf{y}- \bar{\mathbf{H}} \hat{\mathbf{x}} \Vert^2 + \sum_{n=1}^N \Vert \bar{\mathbf{H}}(:.n)\Vert^2 \hat{v}^2_n\right) \nonumber \\&+ \sum_{n=1}^N \sum_{k=1}^K\pi^k_n \log \pi^k_n + \text{constant},\label{eq:ELBO_qj_rewrite}
    \end{align}
    where $q_{\setminus n}(\mathbf{x}_{\setminus n}) = \prod_{n' \neq n} q_{n'}(x_{n'})$.

    To proceed with the optimization, we derive a scalar equivalent observation model for each symbol $x_n$. This involves three main steps: computing the residual signal, deriving the equivalent scalar model, and updating the posterior distribution.

    \begin{algorithm}[t]
        \caption{Variational Bayesian AFDM Detector} 
        \label{alg:vb_detector}
        \textbf{Inputs}: Received signal $\mathbf{y}$, effective channel matrix $\bar{\mathbf{H}}$, noise variance $N_0$, modulation order $K$, max iterations $N_{\text{iter}}$, tolerance $\epsilon$.  
        \\ \textbf{Initialization}: 
            $\boldsymbol{\mu}_n^{(0)} \gets \mathbf{y}$,
            $\boldsymbol{\pi}^{(0)} \gets \mathbf{1}/K$,  
            $\hat{\mathbf{x}}^{(0)} \gets \bar{\mathbf{H}}^H (\bar{\mathbf{H}}\bar{\mathbf{H}}^H + N_0\mathbf{I})^{-1} \mathbf{y}$,  
            $\hat{\mathbf{v}}^{(0)} \gets \mathbf{1}$.
        \\1: \textbf{for} $t = 1$ to $N_{\text{iter}}$ \textbf{do}  
        \\2: \hspace*{3mm}\textbf{for} each symbol $j = 1$ to $N$ \textbf{do}  
        \\ \hspace*{10mm}\textbf{Step 1:} Compute the residual signal.
        \\3: \hspace*{6mm}$\boldsymbol{\mu}_n^{(t)} \gets \mathbf{y} - \sum_{n' \neq n}\mathbf{H}(:,n') \hat{\mathbf{x}}_{n'}^{(t-1)}$  
        \\ \hspace*{10mm}\textbf{Step 2:} Derive the equivalent scalar model.
        \\4: \hspace*{6mm} ${\sigma_n^2}^{(t)} \gets N_0 + \sum_{{n'} \neq n} \|\mathbf{H}(:,{n'})\|^2 \hat{v}_{n'}^{(t-1)}$  
        \\5: \hspace*{6mm} $z_n^{(t)} \gets \frac{\mathbf{H}(:,n)^\mathsf{H}  \boldsymbol{\mu}_n^{(t)}}{\|\mathbf{H}(:,n)\|^2}$  
        \\ \hspace*{10mm}\textbf{Step 3:} Update the posterior distribution.
        \\6: \hspace*{6mm}${\pi_{n}^k}^{(t)} \propto \exp\left(-\frac{|a_k - z_n^{(t)}|^2}{{\sigma_n^2}^{(t)}}\right),\ \forall a_k \in \mathbb{A}$  
        \\7: \hspace*{6mm}$\hat{x}_n^{(t)} \gets \sum_{k=1}^Q {\pi_{n}^k}^{(t)} a_k$,  $\hat{v}_n^{(t)} \gets \sum_{k=1}^Q {\pi_{n}^k}^{(t)} |a_k - \hat{x}_n^{(t)} |^2$  
        \\8: \hspace*{3mm}\textbf{end for}  
        \\9: \hspace*{3mm}\textbf{if} \text{satisfies converge condition} \textbf{then break}  
        \\10: \textbf{end for}  
        \\ \textbf{Output}: Detected symbols $\hat{\mathbf{x}} \gets \arg\max_{\mathbb{A}} \boldsymbol{\pi}$.  
    \end{algorithm}

    \textbf{Residual and Effective Noise Variance:} When updating $q_n(x_n)$, we fix variational parameters $\hat{x}_{n'}$, $\hat{v}_{n'}$ for ${n'} \neq n$. The residual signal $\boldsymbol{\mu}_n$ and the effective noise variance $v_n \in \mathbb{R}_{+}$ are given by
    \begin{align} 
    \boldsymbol{\mu}_n &= \mathbf{y} - \sum_{{n'} \ne n} \mathbf{H}(:,{n'}) \hat{x}_{n'}, \\ v_n &= N_0 + \sum_{{n'} \ne n} \Vert \mathbf{H}(:,{n'}) \Vert^2 \hat{v}_{n'}. 
    \end{align}
        The residual signal $\boldsymbol{\mu}_n$ represents the received signal with the contributions of all symbols except $x_n$ removed, and $v_n$ accounts for the noise variance plus the uncertainty from other symbols.

    \textbf{Equivalent Scalar Observation Model:} To simplify the joint estimation problem, we derive a scalar observation model for each symbol $x_n$. The effective observation $z_n \in \mathbb{C}$ and its variance $\sigma_n^2 \in \mathbb{R}_{+}$ are computed as
    \begin{align}
        z_n &= \frac{\bar{\mathbf{H}}{(:,n)}^\mathsf{H} \boldsymbol{\mu}_n}{\|\bar{\mathbf{H}}{(:,n)}\|^2}, \label{eq:scalar_observation} \\
        \sigma_n^2 &= \frac{v_n}{\|\bar{\mathbf{H}}{(:,n)}\|^2}, \label{eq:scalar_variance}
    \end{align}
    yielding the following equivalent scalar model:
    \begin{align}
        z_n = x_n + \tilde{w}_n, \label{eq:scalar_model}
    \end{align}
    where \(\tilde{w}_n \sim \mathcal{CN}(0, \sigma_n^2)\) is the effective noise.

    \textbf{Posterior Probability Update:} The variational posterior probability $\pi_n^k$ is determined using the likelihood function derived from $z_n$ and the prior:
    \begin{align}
        \pi_n^k \propto \text{exp}\left(-\frac{|z_n-a_k|^2}{\sigma_n^2} \right) p(x_n=a_k).
    \end{align}

    Since a uniform prior is assumed in (\ref{eq:prior}), the posterior simplifies to a normalized softmax form as follows:
    \begin{align} 
     \pi_n^k = \frac{\exp\left( -\frac{|z_n-a_k|^2}{\sigma_n^2} \right)}{\sum_{k'=1}^{K} \exp\left( -\frac{|z_n-a_{k'}|^2}{\sigma_n^2} \right)}. 
     \end{align}

     \textbf{Posterior Mean and Variance Update:} The updated posterior mean and variance for $x_n$ are computed as
    \begin{align} 
    \hat{x}_n &= \sum_{k=1}^{K} \pi_n^k a_k, \\
    \hat{v}_n &= \sum_{k=1}^{K} \pi_n^k |a_k - \hat{x}_n|^2. \end{align}
    This update process is repeated for each symbol $n$ until convergence. The resulting algorithm is both low in complexity and theoretically grounded, with guaranteed convergence due to the convexity of the ELBO in each coordinate update \cite{PRML, Convex_Opt}. The proposed VB detector is specified in \textbf{Algorithm \ref{alg:vb_detector}}.

     \subsection{Complexity Analysis}
     We provide a brief analysis of the computational complexity of the ZF, LMMSE, MPA, and VB detectors. The computational complexity of the ZF detector primarily stems from matrix inversion, with a complexity of $\mathcal{O}(N^3)$. For the LMMSE detector, the matrix $\bar{\mathbf{H}}^\mathsf{H}\bar{\mathbf{H}} + N_0 \mathbf{I}$ is a Hermitian band matrix, which allows for more efficient inversion using matrix decomposition techniques such as LDL or Cholesky decomposition, thereby reducing the complexity to below $\mathcal{O}(N^3)$. In contrast, the MPA and VB detectors exploit the sparsity of the matrix $\bar{\mathbf{H}}$, resulting in complexity that scales linearly with the number of symbols $N$ \cite{VB}. In each iteration, only the marginal means of the remaining symbols need to be updated. The dominant computational complexities are summarized in Table \ref{tab:complexity}.
         
    \begin{table}[h]
        \centering
        \caption{Comparation of dominent complexity.}
        \label{tab:complexity}
        \begin{tabular}{@{}clc@{}}
        \toprule
        \textbf{Algorithm} & 
        \textbf{Complexity} \\ 
        \midrule 
        MAP       & $Q^{N}$\\
        ZF        & $\mathcal{O}(N^3)$                    \\
        LMMSE      & $<\mathcal{O}(N^3)$                   \\
        MP        & $\mathcal{O}(N_\text{iter}QNP^2)$      \\
        VB        & $\mathcal{O}(N_\text{iter}QNP)$        \\
        \bottomrule
        \end{tabular}
    \end{table}

\section{Simulation Results}
    \subsection{Simulation Setup}
    This section illustrates the performance of several detectors mentioned above via $1{,}000$ channel realizations. Each AFDM frame consists of $N = 2^8$ symbols. The carrier frequency is set to $4$ GHz, with a system bandwidth of $100$ MHz. The channel model includes up to $16$ delay taps and $2$ Doppler taps, simulating rich multipath and moderate mobility environments. The channel gain follows a Rayleigh distribution. The number of propagation paths is set to either $3$ or $5$. Quadrature phase-shift keying (QPSK) modulation is employed throughout the simulations. 

    \subsection{Numerical Results}

\begin{figure}[t]
        \centering
        \includegraphics[width=0.85\linewidth]{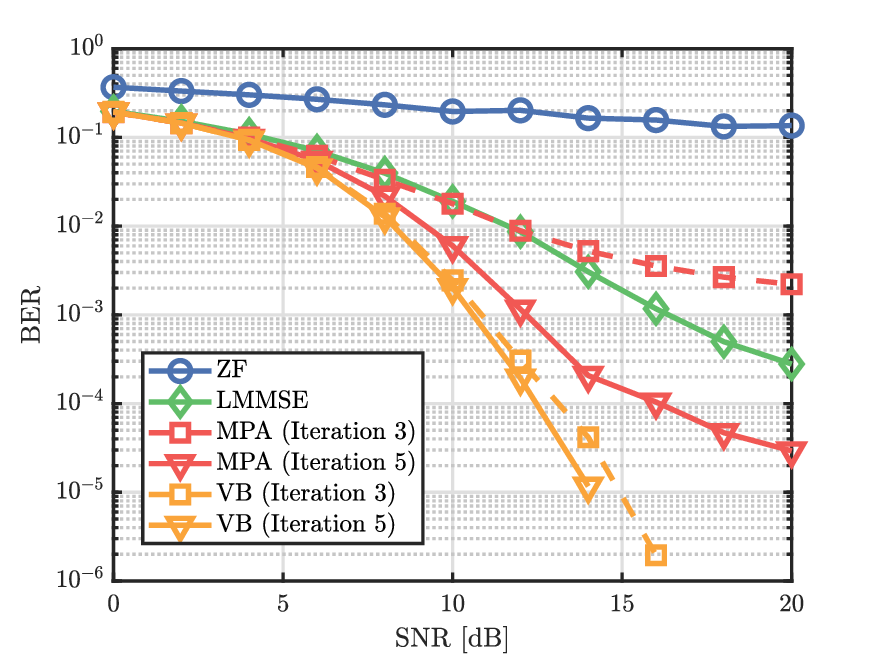}
        \caption{The BER performance for $P=3$ with different algorithms.}
        \label{fig:path3}
    \end{figure}

\begin{figure}[t]
        \centering
        \includegraphics[width=0.85\linewidth]{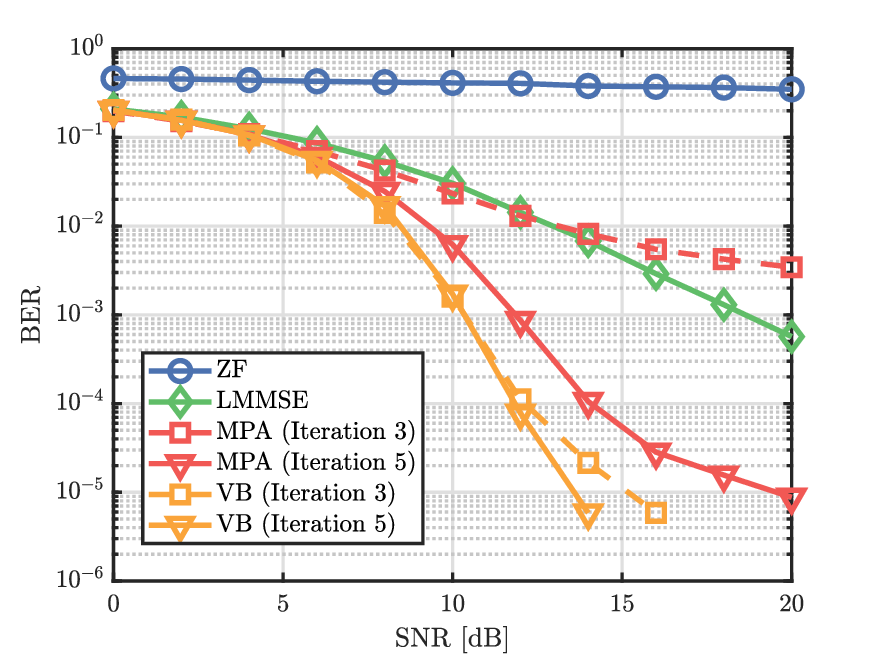}
        \caption{The BER performance for $P=5$ with different algorithms.}
        \label{fig:path5}
    \end{figure}

    Figs.\ref{fig:path3} and \ref{fig:path5} compare the BER performance of detection algorithms under multipath channels with 3 and 5 paths, respectively. Across both scenarios, the ZF detector consistently exhibits the highest BER, showing limited improvement with increasing signal-to-noise ratio (SNR). The LMMSE detector outperforms ZF by balancing noise and interference but saturates at medium-to-high SNR, especially under richer multipath. Nonlinear iterative detectors show clear advantages. With three iterations, MPA surpasses LMMSE across most SNR levels, and five iterations further reduce BER, though gains diminish at high SNR due to message inconsistency in loopy graphs. The VB detector consistently achieves the lowest BER, showing significant improvement with five iterations and a pronounced ``waterfall" curve, especially in complex channels. The performance gap between VB and MPA widens with increased multipath, indicating VB's robustness and scalability in challenging channel conditions.

\begin{figure}[t]
        \centering
        \includegraphics[width=0.85\linewidth]{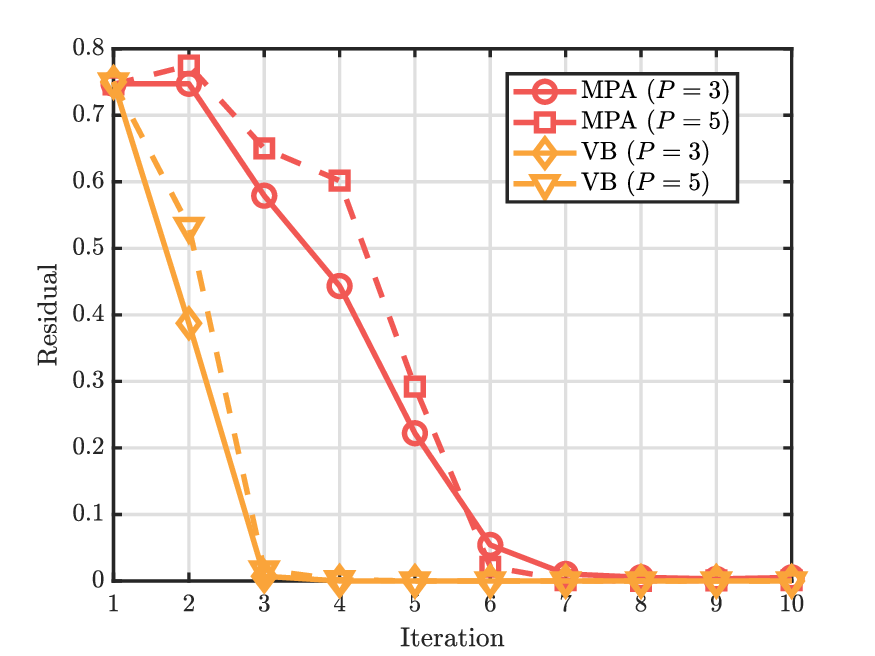}
        \caption{Iteration-wise residuals of MPA and VB detector under different multipath numbers.}
        \label{fig:residual}
    \end{figure}

    The residual curves in Fig. \ref{fig:residual} further explain these trends by showing the per-iteration convergence behavior. The residuals in both MPA and VB detectors quantify the maximum change in probability estimates across iterations to make sure all probabilities have been sufficiently stabilized. For the MPA detector, the residual is defined based on the marginal posterior probabilities $p_i^{(t)}$ of symbol $x_i$ at iteration $t$:
    \begin{align}
        \text{res}_{\text{MPA}}^{(t)} = \max_{n} \left| p_n^{(t)} - p_n^{(t-1)} \right|,
    \end{align}
    where $p_n^{(t)}$ is the edge probability of the $n$-th symbol in the $t$-th iteration. For the VB detector, the residual measures the largest change in posterior probabilities over all transmit symbols and constellation points:
    \begin{align}
        \text{res}_{\text{VB}}^{(t)} = \max_{n,k} \left| {\pi_{n}^k}^{(t)} - {\pi_{n}^k}^{(t-1)} \right|.
    \end{align}
    
    The VB detector converges quickly—typically within three iterations—even with five multipaths, indicating stable posterior estimates early on. In contrast, the MPA detector shows slower convergence and higher early residuals, especially in rich multipath channels, leading to BER saturation. Notably, MPA exhibits residual oscillation at $P=5$. While using maximum residuals helps overall convergence, it may cause large local errors and struggles more as multipaths increase. Overall, VB offers higher accuracy, faster convergence, and better robustness for complex, low-SNR conditions.

\section{Conclusions}
    The proposed VB detector effectively addresses the trade-off between detection performance and computational complexity in AFDM systems. Compared to existing linear detectors and iterative algorithms, the VB detector not only achieves superior BER performance under time-varying channel conditions but also converges rapidly with fewer iterations. Its advantages are particularly pronounced in environments with rich multipath propagation or high SNRs. Due to its fully probabilistic inference framework, the VB detector guarantees theoretical convergence while maintaining low implementation complexity. Consequently, it offers an accurate and efficient detection solution well-suited for practical deployment of AFDM systems in high-mobility wireless communication scenarios. 

\bibliographystyle{IEEEtran}
\bibliography{IEEEabrv, ref}
\balance
\end{document}